# Nanostructuring of $Ba_8Ga_{16}Ge_{30}$ clathrates

Vicente Pacheco[a]*, Raul Cardoso–Gil[b], Deepa Kasinathan[b], Helge Rosner[b], Maik Wagner[b], Lorenzo Tepech–Carrillo[c], Wilder Carrillo–Cabrera[b], Katrin Meier[b] and Yuri Grin[b]

[a]Fraunhofer Institut für Fertigungstechnik und Angewandte Materialforschung, IFAM, Institutsteil Dresden, Winterbergstraße 28, D–01277 Dresden, Germany

[b]Max–Planck–Institut für Chemische Physik fester Stoffe, Nöthnitzer Straße 40, D–01187 Dresden, Germany

[c]Instituto de Física, Universidad Autónoma de Puebla, Apartado Postal J–48, CP–72570 Puebla, México

*Corresponding author. Tel: +49 351 2537 342 Fax: +49 351 2537 399
E–mail: Vicente.Pacheco@ifam–dd.fraunhofer.de

## Abstract

First thermoelectric properties measurements on bulk nanostructured $Ba_8Ga_{16}Ge_{30}$ clathrate-I are presented. A sol-gel-calcination route was developed for preparing amorphous nanosized precursor oxides. The further reduction of the oxides led to quantitative yield of crystalline nanosized $Ba_8Ga_{16}Ge_{30}$ clathrate-I. TEM investigations show the clathrate nanoparticles retain the size and morphology of the precursor oxides. The clathrate nanoparticles contain mainly thin plates (approx. 300 nm x 300 nm x 50 nm) and a small amount of nanospheres (diameter ~ 10 nm). SAED patterns confirm the clathrate-I structure type for both morphologies. The powders were compacted via Spark Plasma Sintering (SPS) to obtain a bulk nano-structured material. The Seebeck coefficient *S*, measured on low-density samples (53% of $\delta_{x\text{-}ray}$), reaches −145 μV/k at 375 °C. The *ZT* values are quite low (0.02) due to the high resistivity of the sample (two orders of magnitude larger than bulk materials) and the low sample density. The trend of the temperature dependence of *S* is in agreement with the values obtained from electronic structure calculations and semi-classical Boltzmann transport theory within the constant scattering approximation. The total thermal conductivity (1.61 W/mK), measured on high density samples (93% of $\delta_{x\text{-}ray}$), shows a reduction of 20-25% in relation to the bulk materials (2.1 W/mK). A further shaping of the sample for the Seebeck coefficient and electrical conductivity measurements was not possible due to the presence of cracks. An improvement on the design of the pressing tools, loading of the sample and profile of the applied pressure will enhance the mechanical stability of the samples. These investigations are now in progress.

## Introduction

The constant increasing demand of energy and the rising cost of energy sources, gas and petrol, are causing a large scientific activity to search for new alternative sources of energy and to render more efficient the processes during transport, conversion and use of energy. Most of these losses occur in the form of wasted heat. The energy harvesting via the recycling and recovery of wasted heat and its conversion into useful electrical energy represents an important approach for facing this energy challenge. Thermoelectric materials can directly convert wasted heat into electricity. The research activities are addressed to develop materials with higher thermoelectric efficiency. The thermoelectric efficiency is related to the so called

*Figure of Merit ZT*= $(S^2\sigma/\kappa)T$, where S, σ and κ are the Seebeck coefficient, the electrical conductivity and the total thermal conductivity. *ZT* of the traditional materials ($Bi_2Te_3/Sb_2Te_3$, PbTe, SiGe) has remained equal or lower than one[1]. Two strategies can be followed to enhance *ZT*, one is increasing the *Power factor* ($S^2\sigma$) and the second is reducing κ. New families of compounds with interesting thermoelectric properties appeared in the last years, skutterudites, silicides, TAGS $(Ge,Te)_{1-x}(AgSbTe_2)_{0.15}$, LAST (PbSbAgTe alloys), Half-Heusler alloys, clathrates. Concerning clathrates, large amounts of studies on the synthesis route, crystal structure and band structure calculations have rise their *ZT* over the historical limit of *ZT* =1. For example; in *n*-type $M_8Ga_{16}Ge_{30}$ clathrates have been reported *ZT* = 1.1 (at 700 K) and *ZT* = 1.4 (at 900K) for M = Eu and M= Ba, respectively [2]. In $Ba_8Ga_{16-x}Al_xSn_{30}$ *ZT* ranges from 0.85 to 1.2 for x=0 and x=6 at 500K, respectively[3]. In *p*-type clathrates, *ZT* > 1 were also found; *ZT* equal to 1.1 and 1.25 for $Ba_8Ga_{16}Ge_{30}$ [2] and $Ba_8Au_{5.3}Ge_{40.3}$,[4] respectively. In 1993, a theoretical work predicted the positive effect of multilayered superlattices (2D)[5] or quantum wires (1D)[6] on the thermoelectric properties. For 3D nanostructured materials, it has been experimentally found that *ZT* could be incremented from 1 to 1.4 at 100 °C in bulk nanostructured BiSbTe alloys.[7] The nanostructuring of the new families of thermoelectric compounds has gained attention among the thermoelectric community.[8-11] The strategy of nanostructuring for further enhancement of *ZT* has not been reported in clathrates. In this work we report the first thermoelectric properties measurements on bulk nanostructured $Ba_8Ga_{16}Ge_{30}$ clathrate-I. A sol-gel-calcination route was developed for preparing amorphous nanosized precursor oxides. The further reduction of the oxides led to quantitative yield of crystalline nanosized $Ba_8Ga_{16}Ge_{30}$ clathrate-I. A long way has been covered in order to find the adequate synthesis conditions of the complete process. The synthesis of the oxides follows an acryl-amide gel route [12] for the production of the precursor oxides. In our first investigations [13], we essentially found the conditions to produce acryl-amide gels containing Eu, Ga and Ge for the production of $Eu_8Ga_{16}Ge_{30}$ clathrates. The proposal of using the sol-gel/calcinations/reduction route for the synthesis of clathrates arise from our previous research where this technique was used to prepare a compound which present chemical similarities, the $Yb_2(Fe,Ga)_{17}$ intermetallic compounds [14] (both systems contain a Rare-earth element and gallium). The synthesis of $Eu_8Ga_{16}Ge_{30}$ clathrates via the reduction of precursor oxides was investigated by L. Tepech [15]. However, the production of $Eu_8Ga_{16}Ge_{30}$ single phase was not possible due to a cascade of phase transformation/decomposition present in the Eu-Ga-Ge system. The route of synthesis we investigate has proved to be successful in the preparation of $Ba_8Ga_{16}Ge_{30}$ clathrate nanoparticles.[16]

## Experimental

### 1.- Synthesis of the $Ba_8Ga_{16}Ge_{30}$ nanoparticles

#### *A. Preparation of the Gel*

BaO, Ga and $GeO_2$ type quartz were used in stoichiometric amounts as source metals for the synthesis of $Ba_8Ga_{16}Ge_{30}$. Ga was dissolved in a mixture of $HNO_3$ concentrated, distillated-demineralized water and citric acid and kept under ultrasonic stirring until the solution turns transparent. BaO was dissolved in the previously prepared $Ga^{3+}$ solution. The amount of citric acid added in the Gallium solution should be enough to complex all the metallic cations (one chemical formula of citric acid per valence of cation). The pH was adjusted to 6.5 using a solution of $NH_4OH$. $GeO_2$ was dissolved in distillated water slightly basic (few drops of $NH_4OH$ concentrated).

All the solutions were mixed together and the volume was completed to 350 ml with distillated-demineralized water. The organic gel was obtained dissolving the monomers of Acrylamide and N N` Methylen bis acrylamide, 3 ml of Hydrogen peroxide solution (3%) was used as initiator of the polymerization reaction. 0.2 ml of TEMED were added as propagator of the polymerization reaction. The solution was placed in a boiling water bath and the polymerization is reached when the water temperature reaches 94 °C to 96 °C.

***Synthesis of Precursor Oxides: Drying and Calcination Process***

The gel is dried at 240 °C under Oxygen flow in an furnace until the liquid phase completely disappears. The dried gel is placed in an alumina boat and calcinated under oxygen flow in a tubular oven for a few hours until the product turns completely white.

#### *B. Calciothermic Reduction*

The as obtained precursor oxides are mixed with a stoichiometric amount of $CaH_2$ and pressed into a pellet. $CaH_2$ will react with the oxygen of the oxides reducing the metals, which will react to form the intermetallic clathrate. The preparation of the Oxides/$CaH_2$ mixture, the pressing into a pellet and the reduction reactions should be performed under argon protective atmosphere. The reduction process takes places in two steps. In a first step a partial reduction occurs at 750 °C and Hydrogen is released. In a second step the complete reduction and synthesis of the intermetallic clathrate takes place at 1130 °C. For the first step, the pellet is placed inside a quartz reactor and next to it a small piece of metallic lanthanum is also positioned for acting as a getter for the traces of moisture and oxygen present in the reactor even after purging several times the atmosphere with argon. The quartz reactor containing the pellet and the La-getter is removed from the glove-box, placed in a tubular oven and connected to a high-vacuum line where the connections should allow the atmosphere could be purged with gettered-Ar. An additional valve is assembled to a bubbler glass filled with mineral oil for allowing the hydrogen evolution that takes place during the reaction at 750 °C and to avoid the overpressure in the quartz reactor. After the hydrogen is completely released, the sample is cooled down. The pellet presents a grey color. For the second step, the partially reduced sample is sealed in a Tantalum ampoule and then in quartz ampoule. All the manipulations are also performed under protective atmosphere. The quartz ampoule is then heated at 1130 °C to complete the reduction reaction and quenched in ice-water. At the end the pellet presents a black color.

### *C. Removing the CaO matrix: Washing of the sample*

The pellet is crunched and then washed in ultrasonic bath with an aqueous EDTA solution, where the pH was adjusted to 6.5 with $NH_4OH$, for removing the CaO matrix where the $Ba_8Ga_{16}Ge_{30}$ nanoparticles are embedded. After removing the CaO, the sample is rinsed several times with distillated/demineralized water and finally with acetone. The final powder is a very fine black powder.

## 2.- Processing of the nano-particles into a bulk nanostructured material

The $Ba_8Ga_{16}Ge_{30}$ nano-powders are loaded in an 8 mm diameter graphite matrix between one upper and one lower cylindrical graphite punches. A Molybdenum foil and cover plates are used to avoid the contact between graphite and nano-powders. The powder are compacted in a Spark Plasma Sintering machine (Dr. Sinter) completely integrated in a glove box filled with argon to avoid contamination or oxidation. Different strategies where investigated to avoid the crystallite growth during the sintering in order to reach the formation of a high dense material which retains the nano-state of the nano-particles.

## 3.- **Thermoelectric properties characterization.**

The thermal diffusivity (TD) was measured on the sintered pellets (diameter = 8mm, thickness ≈ 1mm) using a Netzsch LFA 447 NanoFlash-Xenon Flash thermal diffusivity device in the temperature range from room temperature up to 300 °C. The Heat capacity at constant pressure (Cp) was measured in a Nestzsch DSC 204F1 Phoenix. The total thermal conductivity was calculated according to the realation $\kappa = Cp \bullet TD \bullet \delta$, where $\delta$ is the density of the sample, at room temperature $\delta$ is measured by the Archimedes method.

After the TD measurements, the pellet was cut and shaped with a wire saw into a prismatic form of 7 x 3 x 3 $mm^3$, approximately. The Seebeck coefficient $S$ and the electrical conductivity $\sigma$ were simultaneously measured with an ULVAC ZEM 3 in the range from room temperature up to 300 °C.

4.- **Characterization techniques**

*A.- Powder diffraction*

Powder X-ray diffraction for the phase identification was carried out in a Huber Image Plate Guinier Camera G670 using Cu $K_{\alpha 1}$ radiation [$\lambda$ = 1.540 598 Å, Ge (111) Johansson monochromator.

Powder X-ray diffraction patterns for profile analysis of the diffraction lines and lattice parameter refinement were collected in a STOE STADI MP powder diffractometer in Bragg-Brentano set up using a Zero background sample holder and Lanthanum hexaboride as internal standard, in case of lattice refinement. Profile analysis was performed on the individual reflections with the purpose of determining the peak broadening and to calculate the crystallite size by a Williamson-Hall Plot analysis. Lattice refinement and profile analysis were carried out with the STOE Win XPOW, Software Package for STOE Powder Diffraction System, Version 2.22, STOE & Cie GmbH, Darmstadt, 2007.

*B.- Thermal analysis*

Thermal analysis of the air and moisture sensitive samples were performed in a Netzsch STA 449C apparatus, completely integrated in a Glove-box (MBraun, Garching, $p(O_2)$, $p(H_2O) < 1$ ppm) filled with Argon ($Al_2O_3$ crucibles, thermocouple type S). Temperature calibration was obtained using five melting standards in the range from 97 °C to 1197 °C

*C.- Electron Microscopy*

The qualitative and quantitative determination of the samples chemical composition was done by using Standard less Energy Dispersive X-ray Spectroscopy (EDXS) in a Phillips XL-30 scanning electron microscope with LaB6-cathod.

Transmission electron microscopy investigations were carried out on FEI Technai 20 with an acceleration voltage of 200 kV. The evaluation of the data was done with the Software Digital Micrograph from Gatan Inc.

*D.- Chemical Analysis*

The samples were chemically digested in acid solution in a microwave (MLS-ETHOS plus) to be prepared for the analysis of the metals content. The analysis is performed in a hermetic teflon container with a VARIAN Vista RL (CCD Simultaneous, ICP-AES).

The excitation and ionization of the analyte is produced with inductive coupled Argon plasma attached to a detector of the specific emission lines of the elements.

For the analytical determination of the carbon content, the samples were weighted, mixed with iron and cooper and sealed in zinc capsules. The analysis is carried out in a LECO C-200 equipment. The samples are heated up to 2200 °C to transform the carbon to CO and finally to $CO_2$ with a catalytic reaction with CuO. The $CO_2$ IR active vibrations are detected with a infrared measuring cell detected and the amount of carbon quantified.

The oxygen determination is performed with a LECO TCH600 equipment in a hot extraction-gas carrier process. All the chemical analyses are performed by triplicate and the average value is calculated.

## Results and Discussions

*Gel formation and drying.*

The preparation of the $Ga^{3+}$ solution from metallic gallium and concentrated $HNO_3$/Citric acid is the longest step of the synthesis. After a few hours of ultrasonic agitation the solution turns into a grey/black turbid solution. The addition of citric acid is to form chelating complex with the cations and to avoid the formation of insoluble hydrates. A clean transparent and colorless solution is obtained after 4 or 5 days of ultrasonic agitation. The dissolution of $GeO_2$ in slightly basic water (2 to 4 drops of $NH_4OH$ concentrated is enough) takes a few minutes under magnetically agitation and the final solution is also transparent and colorless. The dissolution of BaO in the previously prepared $Ga^{3+}$ solution is also immediate. The pH of the $Ga^{3+}$ and $Ba^{2+}$ solution should be adjusted to 6.0 - 6.5 before mixing with the $GeO_2$ solution, otherwise a precipitation of $GeO_2$ will occurs.

To minimize the number of parameters to investigate, the concentration of Acrylamide and NN`methylenbisacrylamide was kept constant during all the experiments, 0.06, m/V and 0.005 m/V, respectively. The pH of the solution prior to the polymerization reaction is an important factor. The gel formation with solutions of pH = 6, 7, 8, and 9 was investigated.

The best results were obtained with pH = 6 where a bluish-white, homogeneous, translucent and stable gel was obtained within a few minutes. The drying of the gels takes between 24 to 48 hr depending on the amount of gel to dry. At the end the gel is transformed into a crispy brownish material where the brown color indicates a partial calcination of the citrates.

*Synthesis of the precursor oxides: Calcination.*

Several profiles and calcinations conditions were investigated in order to obtain the nanoparticles of the precursor oxides free of residual calcination products (carbon). Concretely; the heating/cooling rate, the type of atmosphere (air or oxygen flow) and the temperature and time of calcination were investigated. The best results were obtained under the following conditions: Heating (5°C/min) from room temperature up to 350 °C and holding this temperature for 30 minutes. Then continue heating (5 °C/min) up to 780 °C and calcinate for 4 hours. The type of atmosphere during the process plays an important role. An oxygen flow is used during the heating to 350 °C. As soon as the temperature reaches 350 °C the oxygen flow is changed for an air flow. The air flow is preserved during the heating to 780 °C and kept it for 1 hour at 780 °C, and then the oxygen flow is re-established for 3 hours holding the temperature at 780 °C to complete the 4 hours of calcinations. Finally the sample is allowed to freely cool down to room temperature under oxygen flow. The use of air flow and the dwell at 350 °C is to suppress a very exothermic calcination reaction otherwise present at 390 °C. The exothermic reaction causes the sample gets incandescent and produces a sintering and growing of the precursor oxide nanoparticles. Such effect is harmful for the further preparation of the clathrate nanoparticles. The figure 1 shows the X-ray powder diffraction patterns of the samples calcinated at different temperatures.

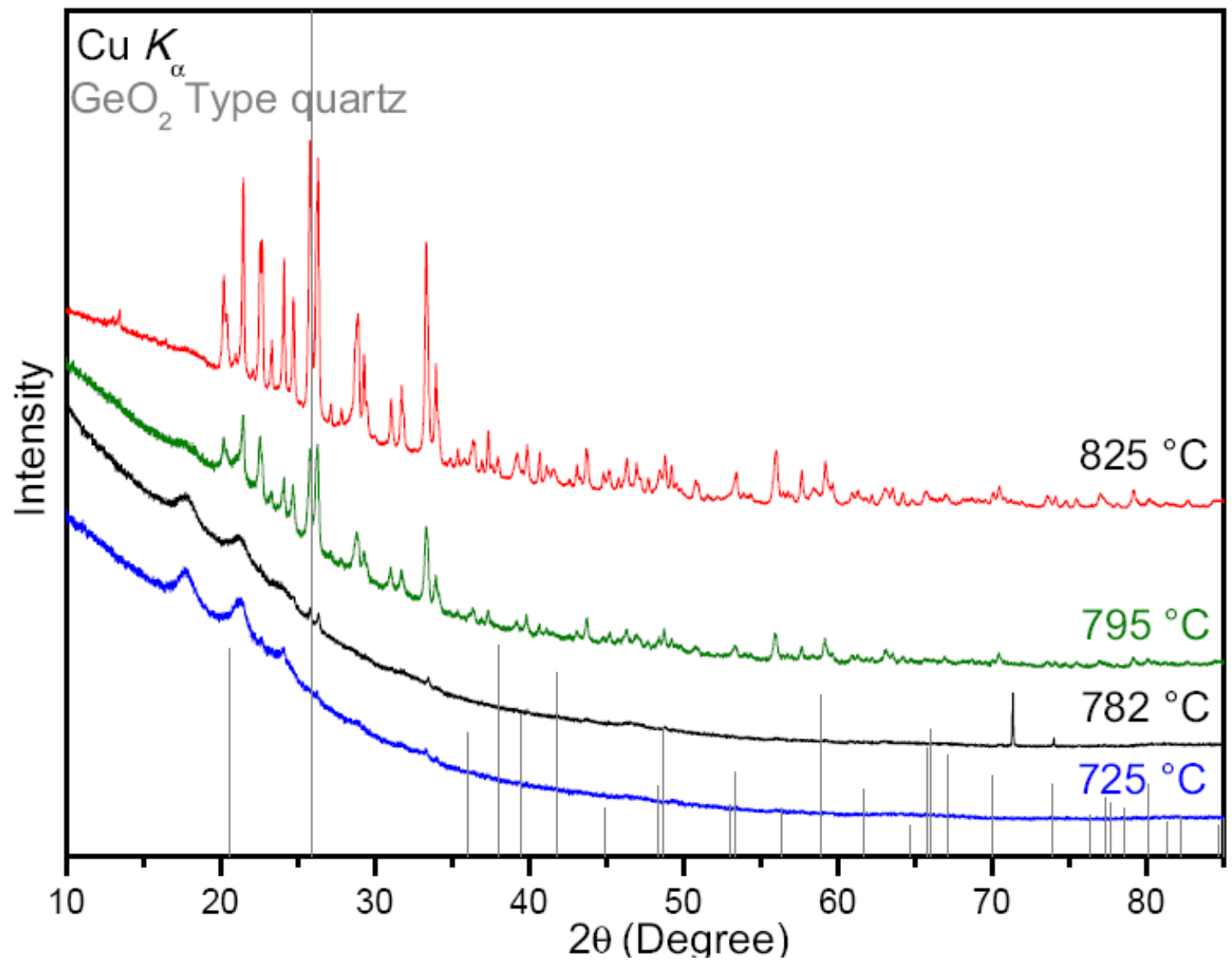

Figure 1. X-ray powder diffraction patterns of the samples calcinated at different temperatures.

A nearly X-ray amorphous sample is obtained for thermal treatments below 725 °C. The two broad diffraction reflections at around 17.5 ° and 21.5° are due to the Vaseline/hexane used to fix the powders on the Kapton foil sample holder of the Guinier camera.

Crystalline phases start to appear when the calcinations are performed at around 780 °C. The amount and the crystallinity of these phases increase as the calcination temperature rise. The crystallization process might also be connected to a crystal growth of the nanoparticles, for this reason it is important to preserve the amorphous state of the precursor oxides. The vertical lines in the powder pattern indicate the reflection positions for $GeO_2$ type quartz. The powder patterns in figure 1 are not fully indexed, and the remaining diffraction lines cannot be attributed to any reported binary or ternary oxide systems. Further investigations on the nature of the non-identified reflections were not performed since it is not the scope of this research to characterize the crystal structure of the precursor oxides.

The carbon content in the samples calcinated at 750 °C and 775 °C is 0.30 ± 0.03 and 0.14 ± 0.01 (in mass %), respectively. For samples calcinated at 780 °C, 785 °C, 805 °C and 825 °C, carbon was not detectable under the detection limit (0.15 %) of the analysis.

The chemical composition of the samples after the optimal calcination is shown in the table 1. The amount of the cations in the precursor oxides is in general in very good agreement with the clathrate composition ($Ba_8Ga_{16}Ge_{30}$). The calculated composition is obtained assuming the maximal oxidation state for each cation, $Ba^{2+}$, $Ga^{3+}$ and $Ge^{4+}$. Only the amount of germanium is slightly lower than the calculated value. This might be due to the volatile character of germanium fluoride formed during the digestions of the sample with hydrofluoric acid (HF) for the analysis.

| Element | Experimental (Mass %) | Calculated (Mass %) |
|---|---|---|
| Ba | 18.59 ± 0.35 | 18.73 |
| Ga | 18.80 ± 0.13 | 19.02 |
| Ge | 34.36 ± 0.51 | 37.15 |
| O | 23.56 ± 0.11 | 25.10 |

Table 1. Experimental and calculated chemical composition of the precursor oxides obtained under optimal calcinations conditions. The nominal cation ratios were set to the clathrate composition ($Ba_8Ga_{16}Ge_{30}$) and the calculated composition was obtained assuming all the cations are in their maximal oxidation state.

*Synthesis of the $Ba_8Ga_{16}Ge_{30}$ nanoparticles by calciothermic reduction*

Since the chemical analysis indicates that each cation in the precursor oxides is present according to the clathrate composition and they are found in their maximal oxidation state then it is valid to propose the following redox chemical reaction just with the aim to quantify the amount of reducing agent ($CaH_2$) it will be required for the synthesis of the intermetallic clathrate:

$$92 \times 2e^-$$

$$8\,\overset{+2}{Ba}O + 8\,\overset{+3}{Ga_2}O_3 + 30\,\overset{+4}{Ge}O_2 + 92\,Ca\overset{-1}{H_2} \longrightarrow \overset{+2}{Ba_8}\overset{-1}{Ga_{16}}\overset{0}{Ge_{30}} + 92\,CaO \quad 92\,\overset{0}{H_2}$$

$$30 \times 4e^-$$

$$16 \times 4e^-$$

According to this chemical equation, the $Ba_8Ga_{16}Ge_{30}$ nanoparticles will be embedded in a CaO matrix and $H_2$ will be released as gas in an amount equivalent to 1.9 in mass %. In spite of the real nature and structure of the precursor oxides are not completely determined; the feasibility of the redox reactions can be anticipated analyzing the existing thermodynamic data for the formation of the binary oxides:

$$\frac{2x}{y}M + O_2 \longrightarrow \frac{2}{y}M_xO_y$$

Where M = Ca, Ba, Ga or Ge. The Gibbs free energy was calculated from the Gibbs–Helmholtz relation, $\Delta_f G° = \Delta_f H° - T\Delta_f S°$, for temperatures up to 1300 °C. The stability of each binary oxide was analyzed by means of an Ellingham Diagram, figure 2. The line corresponding to the $2Ca + O_2 \Rightarrow 2\,CaO$ equilibrium presents the most negative $\Delta_f G°$ value and it is the most stable oxide, therefore metallic calcium can reduce the BaO, $Ga_2O_3$ and $GeO_2$.

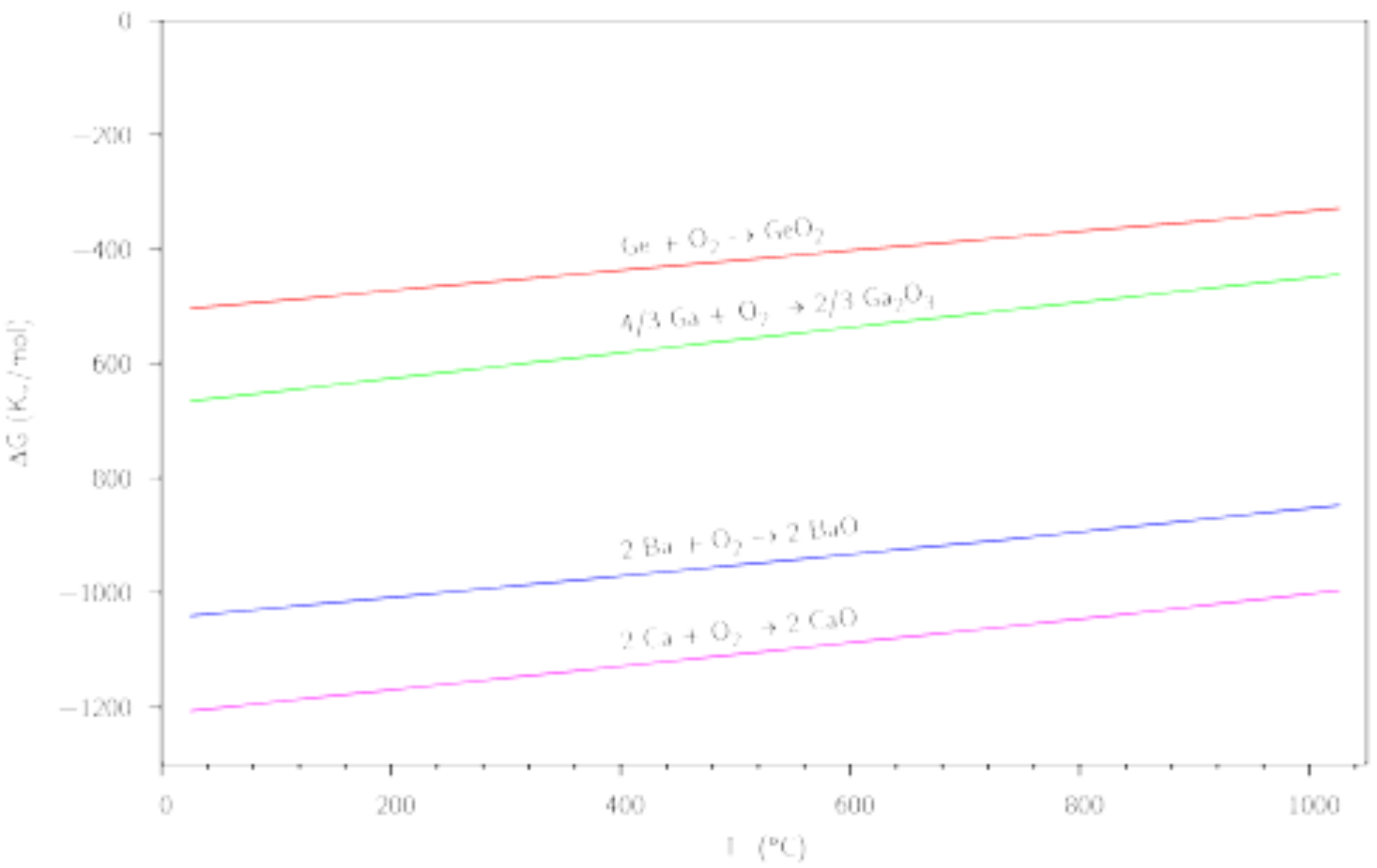

Figure 2. Calculated Ellingham diagrams for the oxides $GeO_2$, $Ga_2O_3$, BaO, CaO

From the figure 2 it can be observed that $GeO_2$ presents the less negative $\Delta_f G°$ value and therefore it will be the first oxide to be reduced. In contrast, BaO is the most stable oxide and it will be the last oxide to be reduced.

The conditions for the chemical reduction of the oxides with $CaH_2$ were investigated *in-situ* by performing DTA-TG experiments, figure 3.

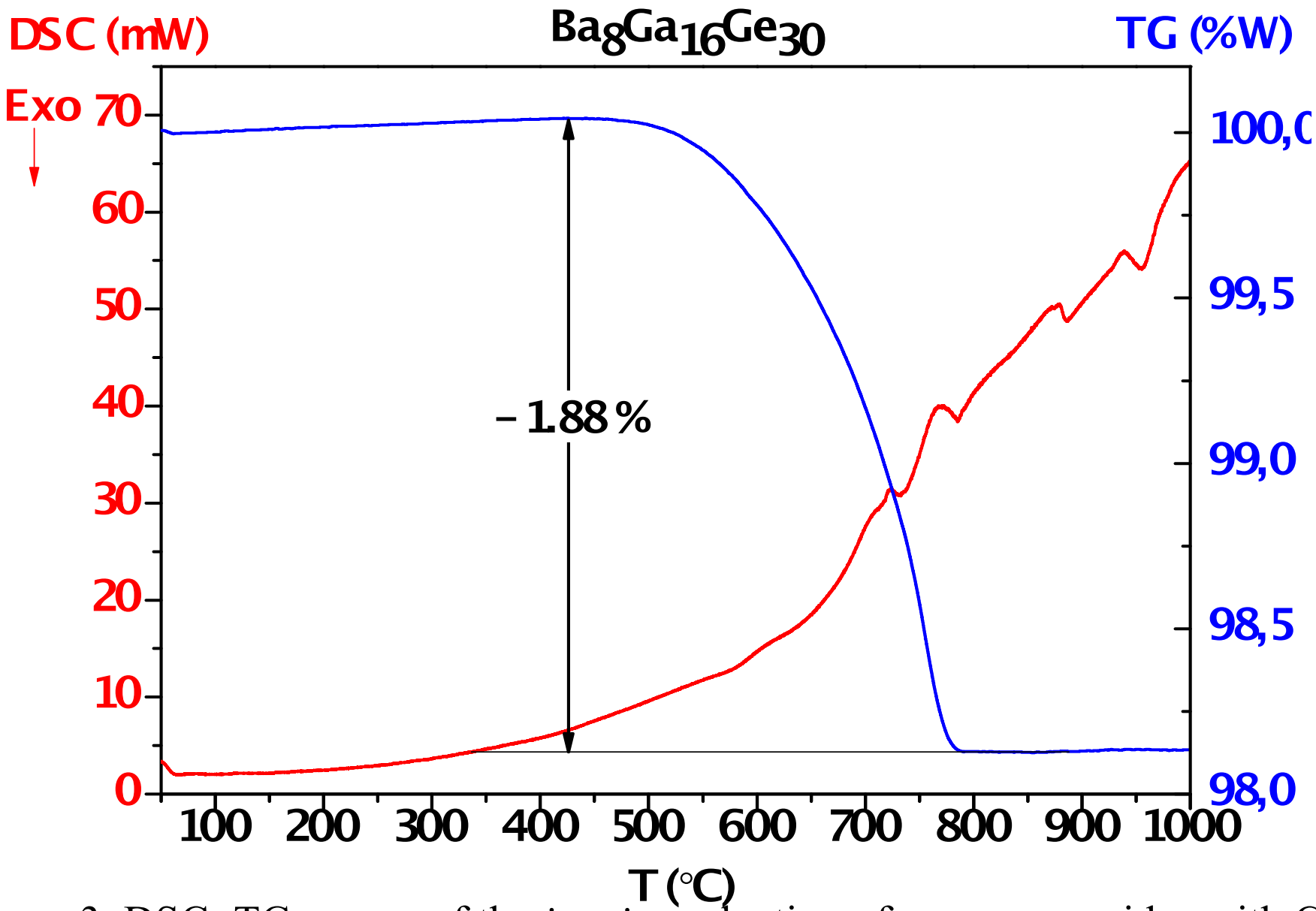

Figure 3. DSC–TG curves of the *in–situ* reduction of precursor oxides with $CaH_2$

The thermogravimetric curve (blue line) shows a mass decrement starting at around 550 °C which extends until approximately 775 °C. After 775 °C, the mass remains constant during the heating up to 1000 °C. The total mass loss is equivalent to 1.88 %, such value is in excellent agreement with the mass loss due to the hydrogen evolution predicted in the proposed redox equation. The complexity of the mechanism of the oxides reduction is evidenced in the DTA curve (red line) which presents a series of thermal effects extending at least up to 1000 °C. The figure 4 shows the powder XRD pattern of a sample reduced at 1130 °C for 30 min and washed with EDTA solution to remove the CaO matrix. Nearly $Ba_8Ga_{16}Ge_{30}$ single phase is obtained as it can be observed by comparing the experimental data (black line) with a calculated powder pattern (vertical red lines). The weak reflection at 27.27° (better seen in the inset) is identified as the strongest (111) reflection of Ge. This is an indication of the presence of Ge traces in the sample.

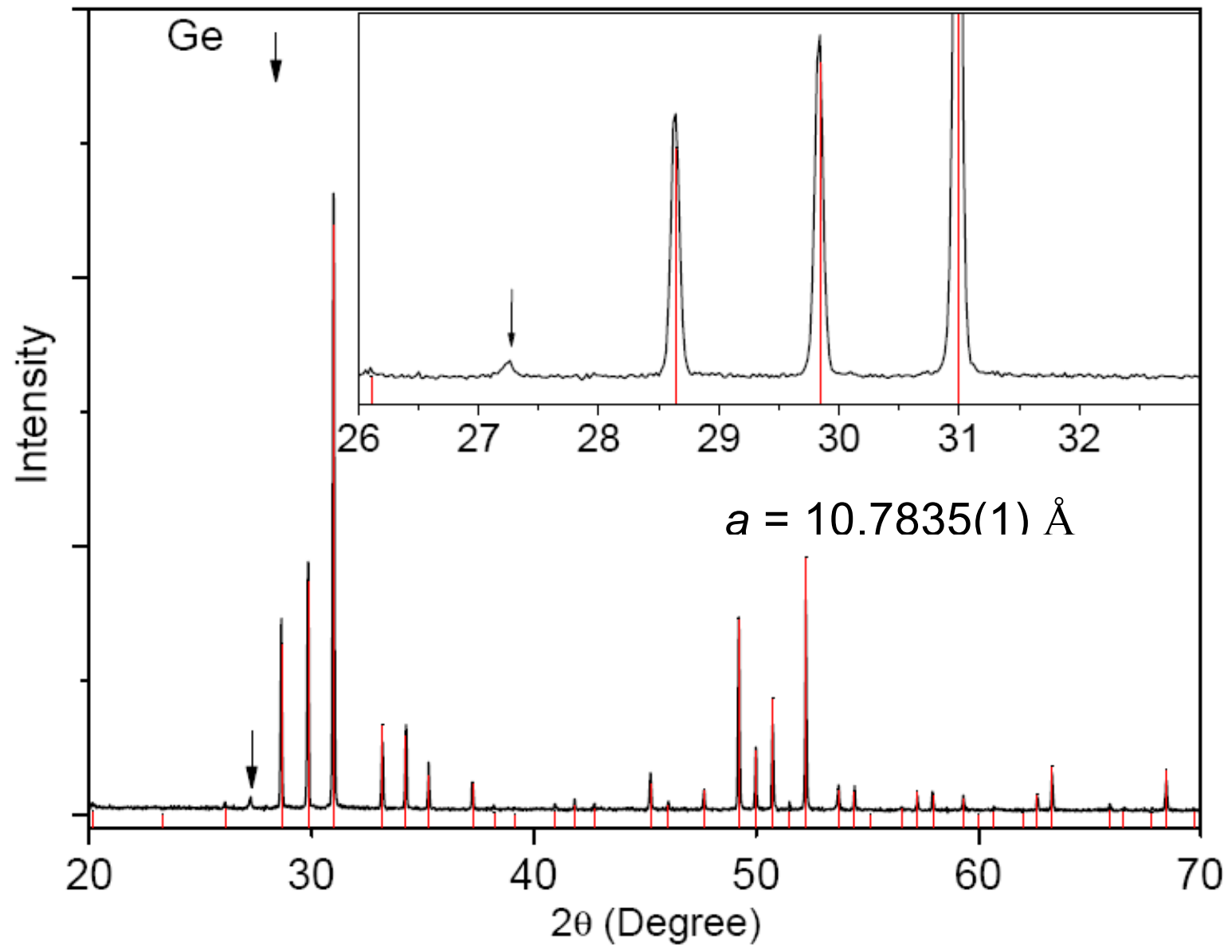

Figure 4. Powder x-ray diffraction patter of the sample reduced at 1130 °C for 30 min. The experimental data (black continuous line) are compared to a calculated pattern (vertical red lines). The arrow marks the strongest (111) reflection of Ge.

*Morphology of the nano-particles.*

As it was presented above, the precursor oxides can be obtained as x–ray amorphous or crystalline materials if the samples were calcined below or above 725 °C, respectively. For the synthesis of the $Ba_8Ga_{16}Ge_{30}$ clathrate nanoparticles the amorphous oxides were used as precursors. The morphology of the particles forming the amorphous precursor oxides is displayed in the figure 5. Two types of particles are recognized. The first types of particles present the form of small conglomerates of maximal 5 μm. A close inspection of the conglomerates show they are formed of interconnected 3D nanoplates. The second types of particles are small conglomerates of nanoparticles. The SAED patterns of the nanoplates present well defined diffraction spots typical of single-crystalline materials. The conglomerates of nanoparticles display very diffuse diffraction rings characteristic of bad crystallized powders or amorphous materials.

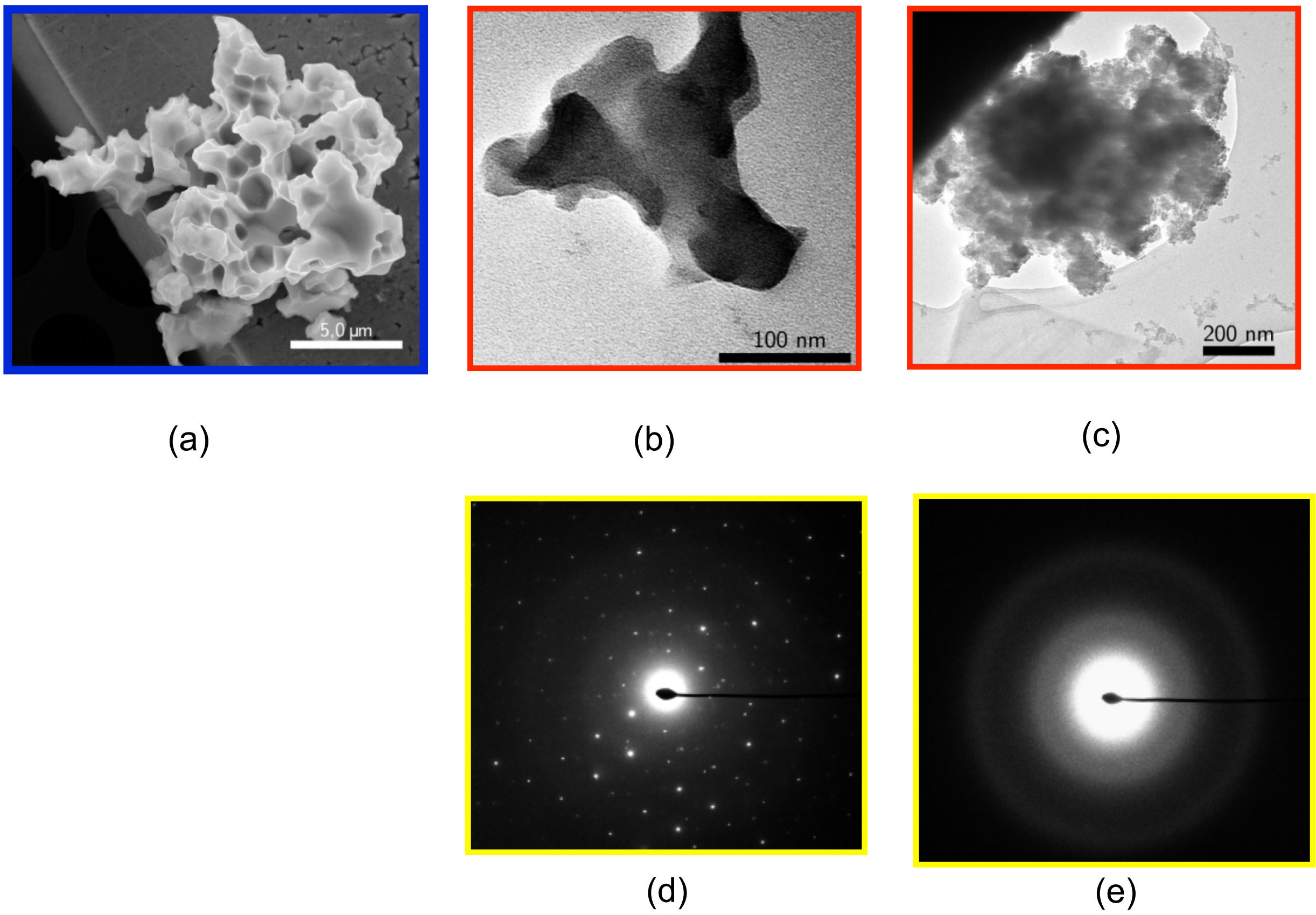

Figure 5. SEM image of the precursor oxides conglomerates (a). The conglomerates are formed by interconnected 3D nanoplates (b). The precursor oxides contain also conglomerates of nanoparticles (c). SAED pattern of the nano-plates (d) and the nanoparticles (e). (a) and (b) are TEM images.

The investigation of the nature and structure of the precursor oxides is not in the scope of this research.

After the reduction of the oxides with $CaH_2$ and once the CaO was removed by washing the sample with EDTA solution, the clathrate nanoparticles are recovered as a very fine powder. The morphology of the particles forming the clatharte nanoparticles is displayed in the figure 6. It is observed that the nanoplates and nanoparticles morphologies of the precursor oxides are retained in the clathrate nanoparticles. The nanoplates size is around 300 nm 300 nm x 50 nm, figure 6a. The SAED pattern of the nanoplates can be indexed according to the [100] direction in the clathrate-I structure type, figure 6b. The lattice parameter of the cubic cell, $a \approx$ 10.9 Å, is in good agreement with the refined lattice parameter, $a \approx$ 10.7835(1) Å, obtained by x-ray powder diffraction. Nanoplates with crystallogtraphic directions other than the [100] direction were also found.

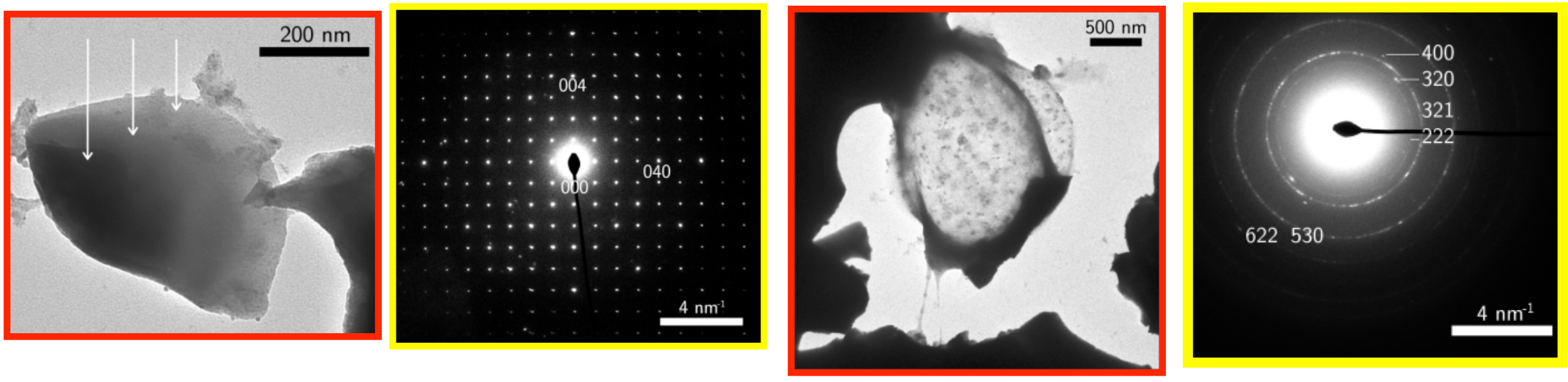

(a) (b) (c) (d)

Figure 6. TEM images and SAED patterns of the $Ba_8Ga_{16}Ge_{30}$ clathrates nanoparticles. As in the precursor oxides, morphologies of nanoplates and nanoparticles were found. The SAED patterns are indexed in the clathrate-I structure type.

Nanoparticles smaller than 50 nm were also found, figure 6c. They represent only a small amount in relation to the nanoplates but no investigations were performed to quantify the relation nanoplates/nanoparticles. It is observed that the nanoparticles are located in a kind of bubble formed in the vortex where 3 or more nanoplates are joining together. In the upper left side of the figure 6c, one complete nanoplate attached to the nanoparticles-bubble is seen. The dark lines in the bubble are rest of nanoplates which should be broken and separated during the processing of the sample. The SAED pattern of the nanoparticles displays diffraction cones characteristic of a crystalline powder. The diffraction cones can be indexed in the clathrate-I structure type according to the (222), (320), (321), (400), (530) and (622) reflections, figure 6d. The discontinuity of the (320) and (400) reflection presume a preferential orientation of the particles or it is simply due to the small amount of particles present in the analyzed volume.

*Sintering of the $Ba_8Ga_{16}Ge_{30}$ clathrates nanoparticles into a bulk nanostructured material.*

The clathrate nanopowders were compacted by Spark Plasma Sintering. In our first experiments a mechanical stable sinter body (8 mm in diameter and 1 mm thick) with low density, 3.13 g/cm$^3$ was obtained. This density represents only the 53.7 % of the x-ray density ($\delta_{x\text{-ray}}$), 5.81 g/cm$^3$. Here the most difficult aspect was the preparation of a sintered material which retains the nano-state of the particles after the sintering. In figure 7a, the powder XRD patterns of the clathrate nanopowders before SPS (figure 4) is compared with the powder pattern after the SPS process. It is shown the material does not present appreciable difference after the compaction and indicates that no structural change was produced. The traces of Ge contained in the sample before the sintering also remain in the sintered specimen. A profile analysis of the diffraction lines was performed in order to calculate the Full Width at the Half of Maximum, (FWHM) and to construct a Williamson-Hall (WH) plot, figure 7b. The WH-plot of the sample before the SPS displays two important information. First; the interception of the least-square-lineal-interpolation (red continuous line) with the ordinate axis results in an average crystallite size of 106 nm. Second, the slope of the linear interpolation (practically horizontal) reveals the powders are free of strain. This is expected since the powders were obtained from bottom-up route and no mechanical stress (milling) was used.

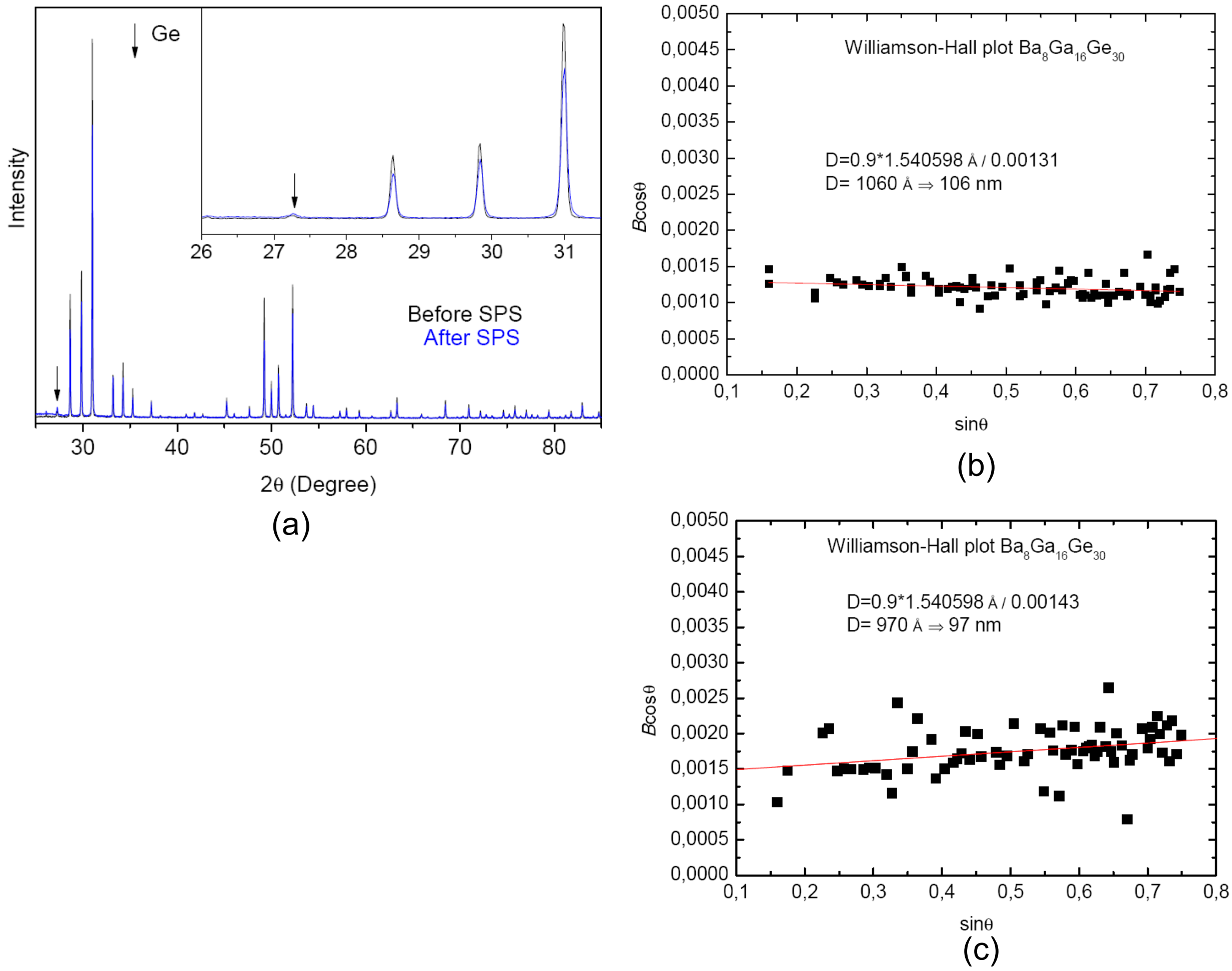

Figure 7. Powder x-ray diffraction patterns of the samples before and after the sintering by SPS (7a). Williamson-Hall plots of the $Ba_8Ga_{16}Ge_{30}$ nano-powders before (7b) and after (7c) the SPS. The red continuous line corresponds to the least square lineal interpolation

The WH-plot of the sample compacted by SPS is shown in the figure 7c. An average crystallite size of 97 nm is obtained from the interception of the least-square-lineal-interpolation (red continuous line) with the ordinate axis. This corroborates that the nanosize of the crystallites were retained after the sintering process. However the data scatter from the linear fit and the slope rise, therefore a small amount of strain was introduced in the sample during the compaction. The preparation of high density bulk nanostructured materials is not a trivial task. The temperature during the compaction process should produce the sintering of the nanoparticles but should avoid the crystallite growth. Try and error were best method to determine the optimal sintering temperature. Once the optimal sintering temperature was found, the applied pressure and holding time were investigate in order to increase the density of the sample. High density samples (94 % of $\delta_{x\text{-}ray}$) were obtained under SPS with optimized parameters 550 °C, 85 MPa, 10 minutes. However, the samples present cracks and further improvements are still in progress to obtain sinter-bodies which could be later cut and shaped for the characterization of the Seebeck coefficient and electrical conductivity.

*Thermoelectric properties of bulk nanostructured $Ba_8Ga_{16}Ge_{30}$ clathrate*

The thermoelectric properties measured in samples with low density (53 % of $\delta_{x\text{-}ray}$) are presented in the figure 8.

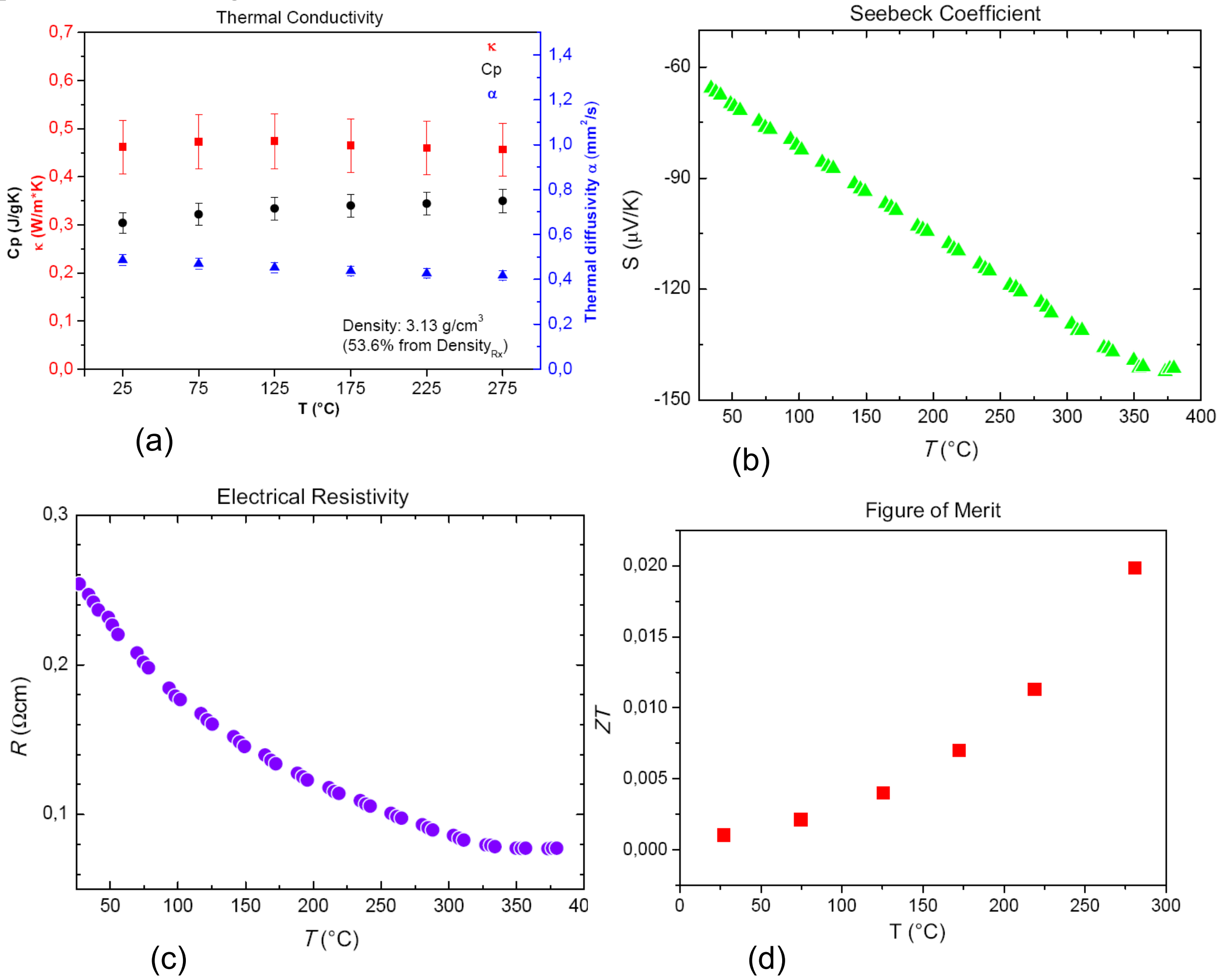

Figure 8. Thermoelectric properties of a low density (53 % of $\delta_{x\text{-}ray}$) bulk nanostructured $Ba_8Ga_{16}Ge_{30}$ clathrate. The numerical value of the left y-axis in the figure 8a is the same for the Cp and the total thermal conductivity, κ. κ was obtained with the relation κ = TD x Cp x δ where TD = thermal diffusivity.

In the figure 8a are shown temperature dependence of the thermal diffusivity (TD), the Cp and the total thermal conductivity (κ).κ is calculated with the relation κ = TD x Cp x δ, where δ is the sample density. In figure 8a the numerical value of the left y-axis is the same for both the Cp and κ. As first approximation, δ was assumed to be constant in the measured temperature range, room temperature up to 275 °C. κ remains practically constant in the measured range and the extremely low value is of κ due to the low density of the sample. The figure 8b shows the temperature dependence of the Seebeck coefficient (S) from room temperature up to 375 °C. The negative value of S indicates the sample is *n*-type and it reaches -145 μV/k at 375 °C. The temperature dependence of the resistivity, figure 8c, indicates the semiconductor character of the sample, however the resistivity values are nearly 2 orders of magnitude larger than the bulk materials. The calculated Figure of Merit *ZT* for the low density sample is presented in the figure 8d. The *ZT* is round 50 times smaller than the best bulk values reported in the literature. The explanation for such a feeble *ZT* is the high resistivity of the sample and therefore the increment of the sample density is mandatory in order to improve the electrical conductivity.

The figure 9a shows the temperature dependence of the thermal diffusivity (TD), the Cp and the total thermal conductivity (κ) for a bulk nansotructured $Ba_8Ga_{16}Ge_{30}$ sample with high density (93 % of $\delta_{x\text{-ray}}$). The numerical value of the left y-axis is the same for both the TD and κ. As in the previous experiment, κ is nearly constant in the measured temperature range. Here was also assumed the sample density is constant from room temperature up to 275 °C. The total thermal conductivity of the high density sample at room temperature (1.61 W/mK) is 3.6 times larger than κ of the low density sample (0.45 W/mK). Nevertheless, this value represents a 20-25 % reduction in comparison to the reported values (2.1 W/mK)[17,18] of bulk materials.

The figure 9b shows the aspect of the high density sintered $Ba_8Ga_{16}Ge_{30}$ pellets after the SPS. The specimens are mechanical stable to perform the characterization of the thermal conductivity but they present cracks. A further shaping of the sample for the Seebeck coefficient and electrical conductivity measurements was not possible. Since the sintering temperature can only be raised until a certain limit to avoid the growth of the crystallites the only possibility to increase the density of the sample is applying higher pressure.

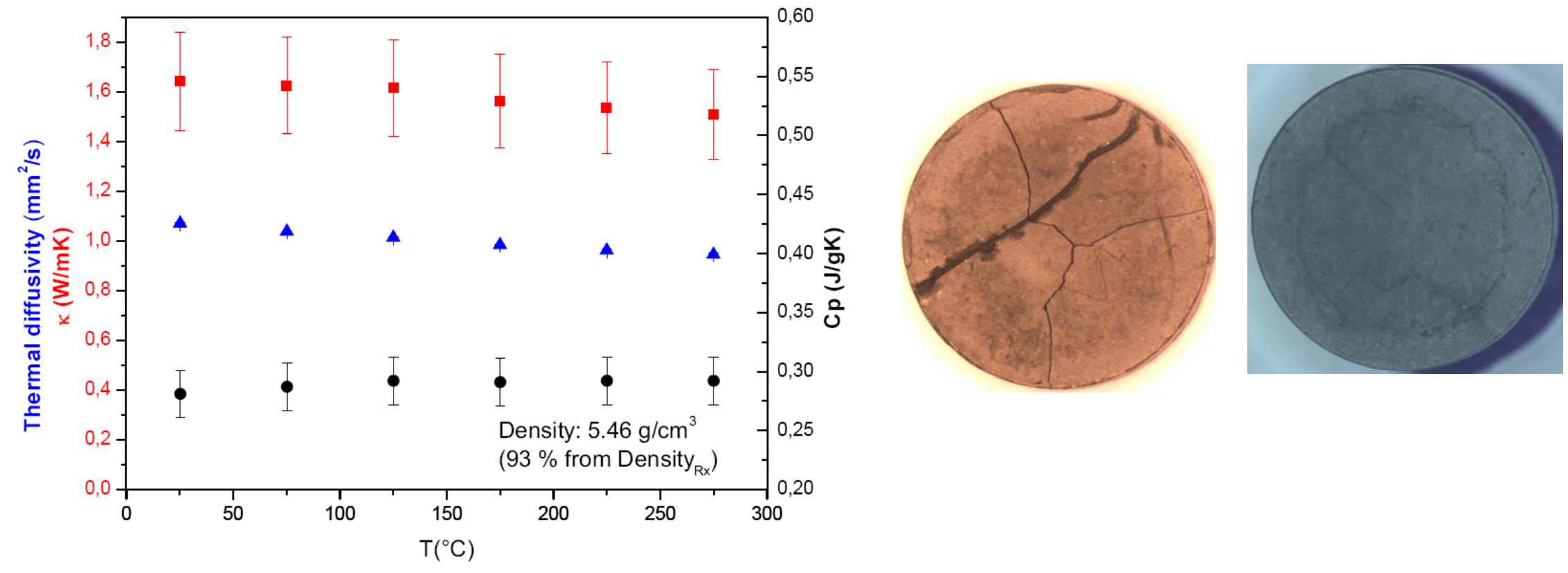

(a) (b)

Figure 9a. Temperature dependence of the thermal diffusivity (TD), Cp and the total thermal conductivity (κ) for a high density (93 % of $\delta_{x\text{-ray}}$) bulk nansotructured $Ba_8Ga_{16}Ge_{30}$, 9a The numerical value of the left y-axis is the same for the TD and κ.

The applied pressure, 85 MPa, is the upper limit that the graphite matrix and punches used in the SPS can tolerate. With this pressure it was possible to reach a density of 5.46 g/cm$^3$ (93% of $\delta_{x\text{-ray}}$). Under such condition, the sample is very sensitive to the non uniform distribution of the applied force. A better loading of the sample produced samples with less cracks, figure 9b right. An improvement on the design of the pressing tools, loading of the sample and profile of the applied pressure will enhance the mechanical stability of the samples. These investigations are now in progress.

*First Principles Calculations of Seebeck Coefficient*

We have calculated the electronic and thermoelectric properties of thin slabs of $Ba_8Ga_{16}Ge_{30}$ with [100] surface termination. Non-spin-polarized total energy and Kohn-Sham band-structure calculations were performed applying the full-potential local-orbital code (version FPLO9.01.35)[19,20], within the local density approximation (LDA). The Perdew and Wang flavor[21] of the exchange correlation potential was chosen for the scalar relativistic calculations. The calculations were carefully converged with respect to the number of *k* points. The transport properties were calculated using the semiclassical Boltzmann transport theory[22,23,24] within the constant scattering approximation as implemented in BoltzTraP.[25] This approximation is based on the assumption that the scattering time τ determining the electrical conductivity does not vary strongly with energy on the scale of *kT*. Additionally, no further assumptions are made on the dependence of τ due to strong doping and temperature. This method has been successfully applied to many narrow band gap materials including clathrates and as well as to oxides.[24,26,27,28] Collected in figure 10 are the Seebeck coefficient for a three, four and five layer slabs with [100] termination as a function of temperature for a carrier concentration of ~ $10^{21}$ electrons. For comparison, the Seebeck coefficient of the bulk is also plotted in the same figure. In contrast to the slabs, increase in thermopower at elevated temperatures for the bulk system is observed for lower carrier concentration of ~ $10^{20}$ electrons. Our results clearly show an increase in Seebeck coefficient values for the geometries with reduced dimensions. This is consistent with the experimental observations mentioned above.

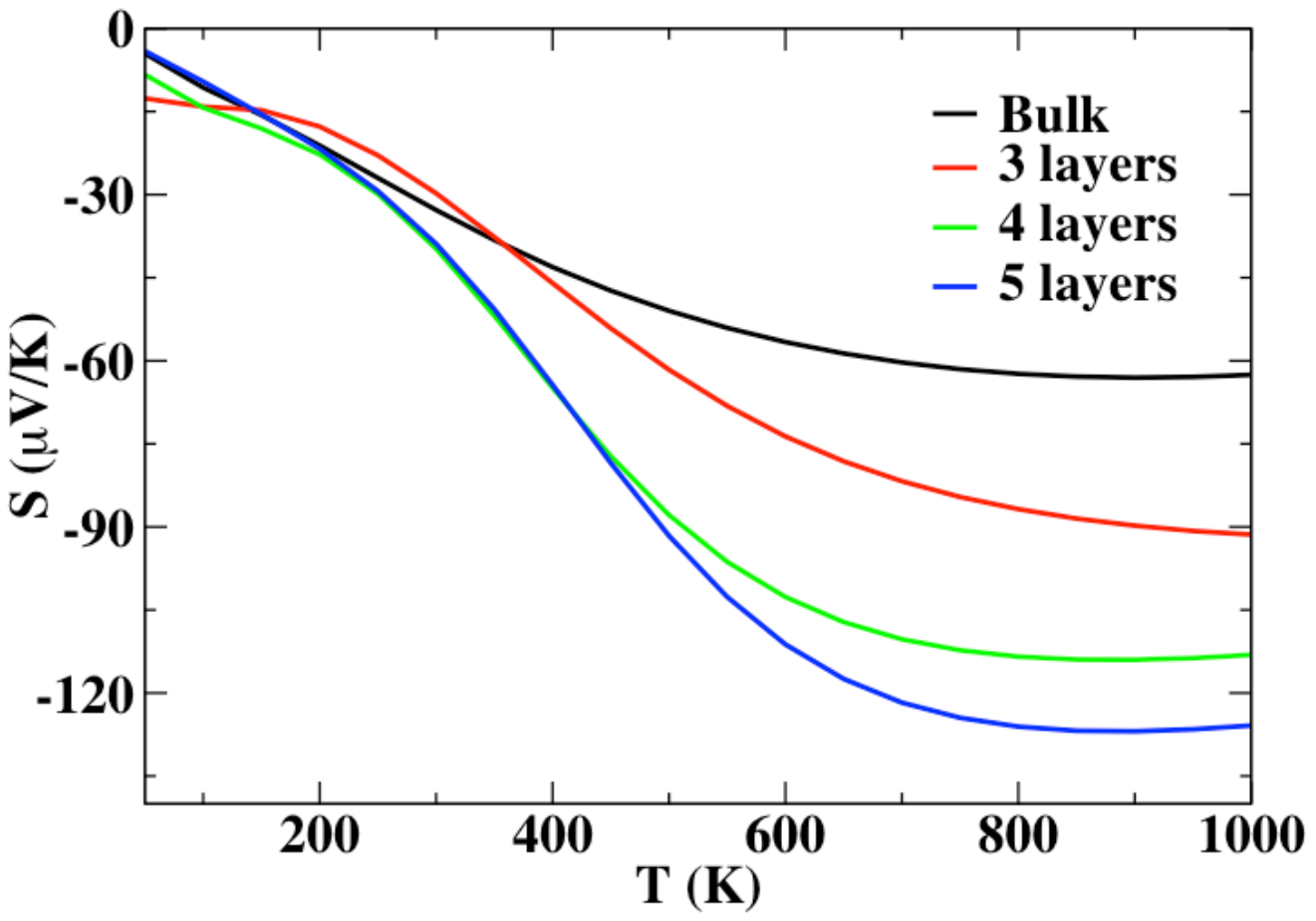

Figure 10. Calculated values of the Seebeck coefficient as a function of temperature for the bulk $Ba_8Ga_{16}Ge_{30}$ system along with three, four and five layer slabs with [100] surface termination. The carrier concentration at which the S increases with temperature is ~$10^{20}$ electrons for the bulk and ~$10^{21}$electrons for the slabs.

*Conclusions*

A sol-gel/calcinations/reduction route was developed for the synthesis of $Ba_8Ga_{16}Ge_{30}$ clathrate nano particles yielding particles with two types of morphologies. The clathrate nanoparticles contain mainly thin plates (approx. 300 nm x 300 nm x 50 nm) and a small amount of conglomerates of nanospheres ($Sphere_{diam.}$ ~ 10 nm). SAED patterns confirm the clathrate-I structure type for both morphologies. The nanoplates are single crystalline and in the results presented here they can be indexed according to the [100] direction. The lattice parameter of the cubic cell, $a \approx 10.9$ Å, is in good agreement with the refined lattice parameter, $a \approx 10.7835(1)$ Å. However, nanoplates with crystallogtraphic directions other than the [100] direction were also found. The nanoparticles displays diffraction cones characteristic of a crystalline powder where the (222), (320), (321), (400), (530) and (622) reflections can be identified. The nanopowders were compacted via Spark Plasma Sintering (SPS) to obtain a bulk nano-structured material. The Seebeck coefficient *S*, measured on low-density samples (53% of $\delta_{x\text{-ray}}$), reaches −145 μV/k at 375 °C. The *ZT* values are quite low (0.02) due to the high resistivity of the sample (two orders of magnitude larger than bulk materials) and the low sample density. The total thermal conductivity (1.61 W/mK), measured on high density samples (93% of $\delta_{x\text{-ray}}$), shows a reduction of 20-25% in relation to the bulk materials (2.1 W/mK).

*Acknowledgments*

This work was supported by the *Deutsche Forschungsgemeinschaft DFG* in the frame of a priority program *SPP1386 "Nanostrukturierte Thermoelektrika: Theorie, Modellsystemme und kontrollierte Synthese"* with the grant number PA 1821/1-1.